\documentclass[twocolumn,showpacs,preprintnumbers,amsmath,amssymb]{revtex4}
\usepackage{graphicx}
\usepackage{dcolumn}
\usepackage{bm}
\usepackage{times}
\usepackage[colorlinks,citecolor=blue,linkcolor=red]{hyperref}
\usepackage{color}
\usepackage{comment}
\usepackage[english]{babel}  

\begin{document}

\title{Two-photon statistics of nonclassical radiation in the dissipative finite-size Dicke model}

\author{Heguang Xu$^{1}$}
\author{Chen Wang$^{1,2}$}\email{wangchenyifang@gmail.com}
\author{Xianlong Gao$^{1,}$}\email{gaoxl@zjnu.edu.cn}
\address{
$^{1}$Department of Physics, Zhejiang Normal University, Jinhua 321004, Zhejiang , P. R. China\\
$^{2}$Shanghai Key Laboratory of Special Artificial Microstructure Materials and Technology,  Tongji University, Shanghai 200092, China
}

\date{\today}

\begin{abstract}
The nonclassical feature of photons in the open finite-size Dicke model is investigated via the two-photon correlation function.
The quantum dressed master equation combined with the extended coherent photonic states is applied to analyze the dissipative dynamics of
both the photons and qubits.
The anti-bunching to bunching transition of photons is clearly observed by tuning the qubit-photon coupling strength.
The optimal qubits number is unraveled to enhance the two-photon correlation function.
Moreover, the temperature bias of thermal baths induces significant two-photon bunching signature with deep strong qubit-photon interaction.

\end{abstract}


\maketitle

\section{Introduction}

The light-matter interaction plays a fundamental role in understanding the optical coherence of quantum theory,
which was originally characterized by R. Glauber~\cite{rjglauber1963pr}.
It has been extensively investigated in quantum optics~\cite{jqyou2011nature,pforndiaz2019rmp}, quantum information processing~\cite{MMariantoni2011science},
quantum dissipation~\cite{uweiss2008book} and quantum materials~\cite{lbritnell2013science}.
The coupling between the radiation field and quantum matter induces the attractive nonclassical feature, exhibiting the effective photon-photon correlation~\cite{aimamoglu1997prl,prabl2011prl,aridolfo2012prl,aridolfo2013prl}.
Based on the theory of quantum photon detection, the statistics of photon nonclassicality can be measured via the intensity correlation function~\cite{prabl2011prl,hjcarmichael2008book}.




One prototype system to describe the quantum light-matter interaction is the quantum Rabi model,
which is composed by a two-level qubit interacting with a single mode radiation field~\cite{rabi1936pr,rabi1937pr,dbraak2016jpa}.
It has been theoretically studied ranging from the quantum optics~\cite{mscully1997book}, quantum entanglement~\cite{qhchen2010pra} to quantum phase transition~\cite{mjhwang2015prl,mjhwang2016prl,mxliu2017prl}.
In particular, the integrability of the Rabi model was recently explored by D. Braak~\cite{dbraak2011prl} and Q. H. Chen~\cite{qhchen2012pra}
with the Bargmann space and extended coherent state approaches, respectively.
The quantum Rabi model was experimentally realized in the cavity-QED platform, with the interaction between photon and qubit reaching the ultrastrong coupling regime (i.e. $\lambda/\omega{\geq}0.1$, $\lambda$ is the coupling strength and $\omega$ the bare frequency of photons).
Accordingly, the traditional rotating-wave-approximation becomes inapplicable.
Another seminal system is the quantum Dicke model, which constitutes of the multi-qubits coupled to a single cavity mode~\cite{rhdicke1954pr,pkirton2019aqt}.
{Besides the transition from the normal phase to the superradiant phase,
which shows the universal scaling behaviors~\cite{nlambert2004prl,qhchen2008pra}, other nonclassical states of light have been investigated via strongly coupled cavity systems~\cite{aimamoglu1997prl,rhuang2018prl,bjli2019pr,
hjcarmichael2015prx,aleboite2016pra,yclu2016qip,hjsnijders2018prl}.}

The representative phenomenon to exhibit the nonclassical character of the radiation field is the photon-blockade effect,
in which the existence of one photon in the cavity strongly suppresses the simultaneous excitation of another photon~\cite{aimamoglu1997prl}.
It is characterized by the dramatic photon antibunching signal.
Such blockade effect has been extensively investigated in various devices, e.g., optomechanical systems~\cite{rhuang2018prl,bjli2019pr}, cavity-QED~\cite{hjcarmichael2015prx,aleboite2016pra,hjsnijders2018prl}
and superconducting circuits~\cite{ajhoffman}.
Particularly for the open quantum Rabi model, it is interesting to find that via the two-photon correlation function the standard photon-blockade breaks down in strong qubit-photon coupling regime~\cite{aridolfo2012prl,aridolfo2013prl}.
A giant photon-photon bunching feature is clearly demonstrated~\cite{aleboite2016pra}.
However, as the multi-qubits analogy, the photon correlations of the Dicke model is preliminarily studied in the quantum phase transition with Kerr nonlinearity~\cite{xyguo2011josb}.
{Due to the finite system size and the availability of the strong coupling regime of the simulated experiments~\cite{baumann2010nature,Sundaresan2019},
the interplay between the finite number of qubits and strong qubit-photon interaction is intriguing to explore.}

In this paper, we study the nonclassical radiation in the dissipative finite-size Dicke model via the two-photon statistics.
The influence of the finite qubits number on the two-photon correlation is investigated, and the transition from the photon anti-bunching to bunching
is clearly exhibited.
Moreover, the optimal enhancement effect is discovered.
The effect of the temperature bias on the two-photon correlation is also analyzed.
It is found that the large temperature bias significantly enhance the photon correlation in strong qubit-photon coupling regime.
The paper is organized as follows:
in section II A, we describe the Dicke model; in section II B and C we apply the quantum master equation combined with the extended coherent photon state to obtain the
dynamics equation of the qubit-photon hybrid system; and in section II D we introduce the two-photon correlation function.
In section III, we study the effects of finite qubits number and finite bath temperatures on the two-photon correlation.
Finally, we give a conclusion in section IV.


\section{Model and method}
\subsection{Dicke model}
The Dicke model, composed by $N$ identical two-level qubits interacting with a single bosonic field, is described as ($\hbar=1$)~\cite{rhdicke1954pr,pkirton2019aqt}
\begin{equation}~\label{h0}
\hat{H}_D={\omega}\hat{a}^{\dagger}\hat{a}+\Delta \hat{J}_{z}+\frac{2\lambda}{\sqrt{N}}(\hat{a}^{\dagger}+\hat{a})\hat{J}_{x},
\end{equation}
where $\hat{J}_{x}=\frac{1}{2}(\hat{J}_{+}+\hat{J}\text{\_})$ and $\hat{J}_{z}$ are the pseudospin operators, composed by $\hat{J}_{{\pm}}=\text{\ensuremath{\sum}}_{i}^{N}\hat{\sigma}_{\pm}^{i}, \hat{J}_{z}=\sum_{i}^{N}\hat{\sigma}_{z}^{i}$, with $\hat{\sigma}_\alpha~(\alpha=x,y,z)$ the Pauli operators and
$\hat{\sigma}_{\pm}=\hat{\sigma}_{x}\pm i \hat{\sigma}_{y}$.
They have the commutating relation
$[\hat{J}_{+},\hat{J}_{-}]=2\hat{J}_{z}$ ,$[\hat{J}_{z},\hat{J}_{\pm}]=\pm \hat{J}_{\pm}$.
$\hat{a}^{\dagger}$ and $\hat{a}$ are the field creating and annihilating operators,
$\Delta$ and $\omega$ are the frequencies of the qubits and single bosonic mode,
and $\lambda$ is the qubit-boson coupling strength.
In the large $N$ limit, the Dicke model undergoes a quantum phase transition~\cite{nlambert2004prl,qhchen2008pra}, where the system transits from the normal phase to the superradiant phase,
with the critical qubit-boson coupling strength $\lambda_c=\sqrt{\omega\Delta}/2$.
While for $N=1$, the Dicke model is reduced to the seminal quantum Rabi model~\cite{rabi1936pr,rabi1937pr}
$\hat{H}_R=\omega \hat{a}^{\dagger}\hat{a}+\frac{\Delta}{2}\hat{\sigma}_{z}+\lambda(\hat{a}+\hat{a}^{\dagger})\hat{\sigma}_x$.

\subsection{Extended coherent bosonic state approach}
The extended coherent bosonic state approach is considered as an efficient method to numerically solve the Dicke model with finite number of qubits~\cite{qhchen2008pra}.
Before including the extended coherent bosonic state method, we first rotate the angular momentum operators with $\pi/2$ along the y-axis
$\hat{H}_0=\exp(i{\pi}\hat{J}_y/2)\hat{H}_D\exp(-i{\pi}\hat{J}_y/2)$, resulting in
\begin{eqnarray}\label{H0bytrans}
\hat{H}_0=\omega \hat{a}^{\dagger}\hat{a}-\frac{\Delta}{2}(\hat{J}_{+}+\hat{J}_{-})+\frac{2\lambda}{\sqrt{N}}(\hat{a}^{\dagger}+\hat{a})\hat{J}_{z}.
\end{eqnarray}
Under the qubits basis $\{|j,m{\rangle},m=-j,-j+1,...,j-1,j$\} with $j=N/2$,
The Hilbert space of the total system can be expressed in terms of
the direct product basis $\{|\varphi_{m}{\rangle}_{b}\otimes|j,m{\rangle}\}$.
In the Dicke model, the excitation number ${N}_{tot}={\langle}\hat{a}^\dag\hat{a}{\rangle}+{\langle}\hat{J}_z{\rangle}+N/2$ is not conserved.
Therefore, the truncation of the bosonic excitation number procedure has to
be applied in this system, especially in the strong qubit-boson coupling regime.
Specifically, by considering the displacement transformation $\hat{A}_{m}=\hat{a}+g_{m}$ with $g_{m}=2\lambda m/\omega\sqrt{N}$
and taking the total system basis into the Schrodinger equation, we obtain
\begin{eqnarray}~\label{eq1}
&&-\Delta j_{m}^{+}|\varphi_{m}{\rangle}_{b}|j,m+1{\rangle}-\Delta j_{m}^{-}|\varphi_{m}{\rangle}_{b}|j,m-1{\rangle}\nonumber\\
&&+\omega(\hat{A}_{m}^{\dagger}\hat{A}_{m}-g_{m}^{2})|\varphi_{m}{\rangle}_{b}|j,m{\rangle}
=E|\varphi_{m}{\rangle}_{b}|j,m{\rangle},
\end{eqnarray}
where $\hat{J}_{\pm}|j,m{\rangle}=j_{m}^{\pm}|j,m\pm1{\rangle}$, with $j_{m}^{\pm}=\sqrt{j(j+1)-m(m\pm1)}$.
Then, we left multiply $\{{\langle}n,j|\}$ to Eq.~(\ref{eq1}), which results in
\begin{equation}
-\Delta j_{n}^{+}|\varphi_{n+1}{\rangle}_{b}-\Delta j_{n}^{-}|\varphi_{n-1}{\rangle}_{b}+\omega(\hat{A}_{n}^{\dagger}\hat{A}_{n}-g_{n}^{2})|\varphi_{n}{\rangle}_{b}=E|\varphi_{n}{\rangle}_{b},
\end{equation}
where $n=-j,-j+1,...,j$.
Furthermore, the bosonic state can be expanded as
\begin{eqnarray}
|\varphi_{m}{\rangle}_{b}&=&\sum_{k=0}^{\textrm{Ntr}}
\frac{1}{\sqrt{k!}}c_{m,k}(\hat{A}_{m}^{\dagger})^{k}
|0{\rangle}_{A_{m}}\nonumber\\
&=&\sum_{k=0}^{\textrm{Ntr}}\frac{1}{\sqrt{k!}}c_{m,k}
(\hat{a}^{\dagger}+g_{m})^{k}e^{-g_{m}\hat{a}^{\dagger}
-g_{m}^{2}/2}|0{\rangle}_{a},
\end{eqnarray}
where $\textrm{N}_{tr}$ is the truncation number of bosonic excitations.
Finally, we obtain the eigen-equation
\begin{eqnarray}
&&\omega(l-g_{n}^{2})c_{n,l}-\Delta j_{n}^{+}\sum_{k=0}^{\textrm{Ntr}}{c_{n+1,k}}_{A_{n}}{\langle}l|k{\rangle}_{A_{n+1}}\nonumber\\
&&-\Delta j_{n}^{-}\sum_{k=0}^{\textrm{Ntr}}{c_{n-1,k}}_{A_{n}}{\langle}l|k{\rangle}_{A_{n-1}}=Ec_{n,l}
\end{eqnarray}
where the coefficients are $_{A_{n}}{\langle}l|k{\rangle}_{A_{n-1}}=(-1)^{l}D_{l,k}$ and
$_{A_{n}}{\langle}l|k{\rangle}_{A_{n+1}}=(-1)^{k}D_{l,k}$,
with
\begin{eqnarray}
D_{l,k}=e^{-G^{2}/2}\sum_{r=0}^{\min[l,k]}\frac{(-1)^{-r}\sqrt{l!k!}G^{l+k-2r}}{(l-r)!(k-r)!r!},G=\frac{2\lambda}{\omega\sqrt{N}}.
\end{eqnarray}

Once we efficiently solve the eigensolution $\hat{H}_0|\phi_k{\rangle}_0=E_k|\phi_k{\rangle}_0$, the original solution can be straightforwardly obtained as
\begin{eqnarray}
\hat{H}_D|\phi_k{\rangle}=E_k|\phi_k{\rangle},
\end{eqnarray}
with $|\phi_k{\rangle}=\exp(-i\pi\hat{J}_y/2)|\phi_k{\rangle}_0$.
For the previous work in analysis of the ground state phase transition with extended coherent bosonic states,
it is surprisingly found that $\textrm{N}_{tr}=6$ is accurate enough to obtain the ground state energy with large qubits number $N=32$~\cite{qhchen2008pra}.
In the following work, we select the truncation number $\textrm{N}_{tr}=50$ up to the $N=160$.

\subsection{Quantum dressed master equation}
For practical light-matter coupled systems, it is inevitable to interact with the dissipative environment, which leads to the Hamiltonian system we studied,
\[
\hat{H}=\hat{H}_D+\hat{H}_B+\hat{V}.
\]
Here, $\hat{H}_0$ is given by Eq.~(\ref{H0bytrans}) and the thermal baths are expressed as,
\[
\hat{H}_B=\sum_{u=q,c}\sum_{k}\omega_k\hat{b}^\dag_{u,k}\hat{b}_{u,k},
\]
where $\hat{b}^\dag_{u,k}~(\hat{b}_{u,k})$ creates (annihilates) one phonon in the $u$th bath with the frequency $\omega_k$.
And the interactions between the Dicke system with thermal baths are specified as
\[
\hat{V}=\hat{V}_q+\hat{V}_c,
\]
with
\begin{eqnarray}
\hat{V}_q&=&\sum_{k}(\lambda_{q,k}\hat{b}^\dag_{q,k}+\lambda^{*}_{q,k}\hat{b}_{q,k}){(\hat{J}_++\hat{J}_-)}/{\sqrt{N}},\\
\hat{V}_c&=&\sum_{k}(\lambda_{c,k}\hat{b}^\dag_{c,k}+\lambda^{*}_{c,k}\hat{b}_{c,k})(\hat{a}^\dag+\hat{a}),
\end{eqnarray}
with $\lambda_{q,k}~(\lambda_{c,k})$ the coupling strength between the qubits (photon) and the corresponding bath. {The $u$th thermal bath is characterized by the spectral function
$\gamma_u(\omega)=2\pi\sum_k|\lambda_{k,u}|^2\delta(\omega-\omega_k)$.
In this paper, we specify $\gamma_u(\omega)$ the Ohmic case $\gamma_u(\omega)=\pi\alpha{\omega}\exp(-|\omega|/\omega_c)$~\cite{uweiss2008book},
where $\alpha$ is the coupling strength and $\omega_c$ is the cutoff frequency of thermal baths.}

By assuming the weak interaction between the Dicke system and thermal baths,
under the Born-Markov approximation,
we obtain the quantum dressed master equation to investigate the dissipative dynamics of the Dicke system as~\cite{aleboite2016pra,fbeaudoin2011pra}
\begin{eqnarray}
\frac{d}{dt}\hat{\rho}_s&=&-i[\hat{H}_0,\hat{\rho}_s]+\sum_{u;k<j}
\{\Gamma^{jk}_un_u(\Delta_{jk})\mathcal{D}[|\phi_j{\rangle}{\langle}\phi_k|,\hat{\rho}_s]\nonumber\\
&&+\Gamma^{jk}_u[1+n_u(\Delta_{jk})]\mathcal{D}[|\phi_k{\rangle}{\langle}\phi_j|,\hat{\rho}_s]\}
\end{eqnarray}
where $|\phi_k{\rangle}$ is the eigenfunction of the Dicke model $\hat{H}_D$ as $\hat{H}_D|\phi_k{\rangle}=E_k|\phi_k{\rangle}$,
the dissipator is
$\mathcal{{D}}[\hat{O},\hat{\rho}_s]=\frac{1}{2}[2\hat{O}\hat{\rho}_s\hat{O}^{\dag}-\hat{\rho}_s\hat{O}^{\dag}\hat{O}-\hat{O}^{\dag}\hat{O}\hat{\rho}_s]$,
the rate is
$\Gamma^{jk}_u=\gamma_u(\Delta_{jk})|S^{jk}_u|^2$,
with ${S}^{jk}_q=\frac{1}{\sqrt{N}}{\langle}\phi_j|(\hat{J}_++\hat{J}_-)|\phi_k{\rangle}$
and ${S}^{jk}_c={\langle}\phi_j|(\hat{a}^{\dag}+\hat{a})|\phi_k{\rangle}$.
In the eigen-basis, the population dynamics is given by
\begin{eqnarray}
\frac{d}{dt}{\rho}_{nn}&=&\sum_{u,k{\neq}n}\Gamma^{nk}_un_u(\Delta_{nk}){\rho}_{kk}\nonumber\\
&&-\sum_{u,k{\neq}n}\Gamma^{nk}_u[1+n_u(\Delta_{nk})]{\rho}_{nn}
\end{eqnarray}
where $\Gamma^{nk}_u=-\Gamma^{kn}_u$.
As $T_q=T_c=T$, the Dicke system at steady state is in thermal equilibrium, such that the equilibrium density matrix operator is ~\cite{aridolfo2013prl}
\begin{eqnarray}
\hat{\rho}_{s}=\sum_k\frac{e^{-E_k/(k_BT)}}{\mathcal{Z}}|\phi_k{\rangle}{\langle}\phi_k|,
\end{eqnarray}
with the partition function $\mathcal{Z}=\sum_ke^{-E_k/(k_BT)}$.
And the steady state population is specified as
\begin{eqnarray}
P_k=e^{-E_k/(k_BT)}/\mathcal{Z}.
\end{eqnarray}

It should be noted that the traditional treatment of the light-matter interacting systems is to apply the Lindblad master equation,
which is proper by considering the weak light-matter interaction.
However, as the light-matter coupling strength becomes strong, the Lindblad equation breaks down.
The dissipative dynamics of the quantum system is suggested to investigate in the dressed picture~\cite{fbeaudoin2011pra},
which makes the transitions between the eigenstates of $H_0$ at Eq.~(\ref{h0}).

\subsection{Zero-time delay second-order correlation function}
In quantum optics, the traditional definition of steady state two-photon correlation function, which was initially proposed by the R. J. Glauber,
is expressed as~\cite{rjglauber1963pr}
\begin{eqnarray}~\label{g2}
G^{(2)}(0)=\frac{{\langle}\hat{a}^{\dag}\hat{a}^{\dag}\hat{a}\hat{a}{\rangle}}{{\langle}\hat{a}^{\dag}\hat{a}{\rangle}^2},
\end{eqnarray}
where ${\langle}\cdots{\rangle}$ means the expectation value at steady state.
$G^{(2)}(0)$ describes the probability of detecting two photons simultaneously,
which is normalized by the probability of detecting two photons at once within a random photon source.
It is known that the bunching and antibunching are two significantly distinguishable features of photon statistics.
Specifically, the bunching(also termed as super-Poisson statistics) dictates that photons populate themselves together,
whereas the antibunching(also termed as sub-Poisson statistics) is the opposite behavior, in which photons distribute separately.
Hence, the antibunching indicates the anticorrelation effect as the second photon is measured.
Quantitatively, the second-order correlation function with the bunching is characterized as~\cite{hjcarmichael2008book}
\begin{eqnarray}
G^{(2)}(0)>1.
\end{eqnarray}
In contrast, the photon antibunching is defined as
\begin{eqnarray}
G^{(2)}(0)<1.
\end{eqnarray}
Moreover, for the thermal state, the correlation function is $G^{(2)}(0)=2$~\cite{hjcarmichael2008book,rjglauber2006rmp}.
Such definition of the two-photon correlation function may be properly applied to investigate photon statistics in Lindblad form open quantum systems with weak light-matter interaction.

However, as the light-matter interaction becomes strong, the two-photon correlation function should be measurement in the eigenbasis.
The normalized and generalized two-photon correlation function of the finite size Dicke model is given by~\cite{prabl2011prl,aridolfo2012prl}
\begin{eqnarray}~\label{gn2}
G^{(2)}_N(0)=\frac{{\langle}(\hat{X}^-)^2(\hat{X}^+)^2{\rangle}}{{\langle}\hat{X}^-\hat{X}^+{\rangle}^2},
\end{eqnarray}
where $N$ is the qubits number, ${\langle}\hat{O}{\rangle}=\textrm{Tr}\{\hat{O}\hat{\rho}_s(t{\rightarrow}\infty)\}$,
\begin{eqnarray}~\label{xp}
\hat{X}^+=-i\sum_{k>j}\Delta_{kj}X_{jk}|\phi_j{\rangle}{\langle}\phi_k|,
\end{eqnarray}
with $\hat{X}^-=(\hat{X}^+)^{\dag}$, $\Delta_{kj}=E_k-E_j$,
and
$X_{jk}={\langle}\phi_j|(\hat{a}^\dag+\hat{a})|\phi_k{\rangle}$.
${X}^+_{jk}$ describes the transition from the higher eigenstate $|\phi_k{\rangle}$ to the lower one $|\phi_j{\rangle}$.
It should be noted that $\hat{X}^+|\phi_0{\rangle}=0$ for the ground state of $\hat{H}_s=\hat{H}_0+\hat{H}_B$, in contrast to $\hat{a}|\phi_0{\rangle}{\neq}0$.
Moreover, in the weak qubit-photon interaction limit (i.e. $\lambda{\approx} 0$),
the operator $\hat{X}^+$ is simplified to $\hat{X}^+=-i\omega\hat{a}$.
Hence, two-photon correlation function in Eq.~(\ref{gn2}) returns back to the counterpart in Eq.~(\ref{g2}).
The expression of correlation function  in Eq.~(\ref{gn2}) has been extensively analyzed in the dissipative quantum Rabi model and optomechanical systems~\cite{aridolfo2012prl,aridolfo2013prl,prabl2011prl}.
In the following, we apply $G^{(2)}_N(0)$ to study the steady state two-photon statistics in the finite qubits number dissipative Dicke model.

\section{Results and Discussions}

\begin{figure}[tbp]
\begin{center}
\includegraphics[scale=0.4]{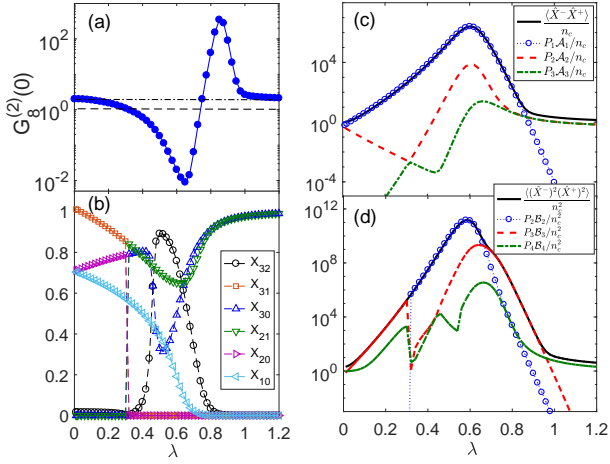}
\end{center}
\caption{(Color online)
(a) Steady state two-photon correlation function $G^{(2)}_N(0)$, with the black dashed line indicating $G^{(2)}_8=1$
and the black dashed-dotted line describing $G^{(2)}_8=2$;
(b) the element of transition operator $X_{jk}$ at Eq.~(\ref{xp});
(c) renormalized one-photon correlation function ${\langle}\hat{X}^-\hat{X}^+{\rangle}/n_c$ and components $P_k\mathcal{A}_k/n_c$
with the Bose-Einstein distribution function $n_c=1/[\exp(\omega/k_BT_c)-1]$;
(d) correlation function ${\langle}(\hat{X}^-)^2(\hat{X}^+)^2{\rangle}/n^2_c$ and components $P_k\mathcal{B}_k/n^2_c$.
The other system parameters are given by
$\Delta=1$, $\omega=1$, $N=8$, $\alpha=0.001$, $\omega_c=10$, and $T_c=T_q=0.05$.
}~\label{fig2}
\end{figure}

\begin{figure}[tbp]
\begin{center}
\includegraphics[scale=0.4]{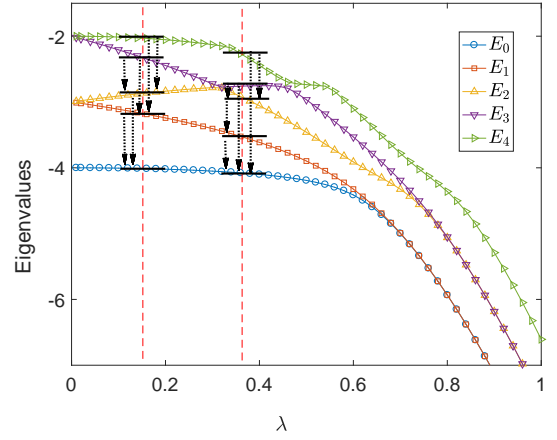}
\end{center}
\caption{(Color online) The five lowest eigenvalues $E_k$ as a function of qubit-photon coupling strength $\lambda$.
Two vertical dashed red lines specify the qubit-photon coupling strengthes as $\lambda=0.15$ and $\lambda=0.35$, respectively;
the horizontal solid black lines describe corresponding eigenstates; and the vertical dashed black lines with arrows shows the transition
between different eigenstates.
The system parameters are the same as in Fig.~\ref{fig2}.
}~\label{fig2-1}
\end{figure}

\begin{figure}[tbp]
\begin{center}
\includegraphics[scale=0.35]{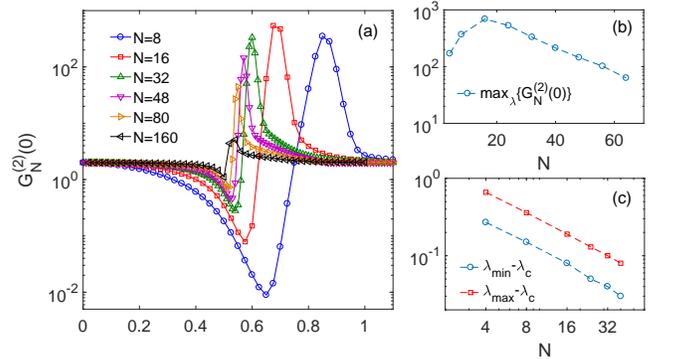}
\vspace{-1.0cm}
\end{center}
\caption{(Color online) (a) Two-photon correlation function $G^{(2)}_N(0)$ as a function of qubit-photon coupling strength $\lambda$ with various qubits number $N$;
(b) the maximum of the two-photon correlation function
$\max_{\lambda}\{G^{(2)}_N(0)\}$ as a function of the qubits number by tuning $\lambda$;
(c) the scaling behavior of the coupling strength bias $\lambda_{\textrm{min}(\textrm{max})}-\lambda_c$,
with $\lambda_{\textrm{min}(\textrm{max})}$ corresponding to the minimum (maximum) of $G^{(2)}_N(0)$,
and $\lambda_c=\frac{\sqrt{\omega\Delta}}{2}\sqrt{\coth(\frac{\Delta}{4k_BT})}$ (see Ref.~\cite{vnpopov1982tmp}).
The other system parameters are given by
$\Delta=1$, $\omega=1$, $\alpha=0.001$, $\omega_c=10$, and $T_c=T_q=0.05$.
}~\label{fig3}
\end{figure}


\subsection{Effect of qubit-photon coupling strength}

We first investigate the effect of qubit-photon interaction on the zero-time delay two-photon correlation function
$G^{(2)}_N(0)$ with $N=8$ in the low temperature regime(e.g., $T_c=T_q=T=0.05\omega$) in Fig.~\ref{fig2} (a).
In the qubit-photon coupling regime $\lambda{\in} (0,0.3)$, the finite eigenenergy difference(see Fig.~\ref{fig2-1}) results in
$P_1{\gg}P_2{\gg}P_3{\gg}P_4$.
The transition between eigenstates $|\phi_2{\rangle}$ and $|\phi_1{\rangle}$ assisted by  thermal baths is prohibited ($X_{21}=0$) due to the same odd parity ${\langle}{e^{i\pi(\hat{a}^\dagger\hat{a}+\hat{J}_z+N/2)}}{\rangle}=-1$, which is schematically shown in Fig.~\ref{fig2-1}.
Moreover, From Fig.~\ref{fig2} (b) it is known $X_{32}{\ll}X_{31}$.
Hence, the two-photon correlation function is simplified by the dominant terms as
\begin{eqnarray}
G^{(2)}_8(0){\approx}P_3(\Delta_{31}X_{31})^2/[P^2_1(\Delta_{10}X_{10})^2].
\end{eqnarray}
It is found that by enhancing the interaction strength $\lambda$,
the two-photon correlation function shows subthermal behavior (i.e. $G^{(2)}_8(0)<2$),
which is the signature of the nonclassical feature.

In the regime $\lambda{\in}(0.3,0.6)$, due to the avoid-crossing of the energy levels $E_2$ and $E_3$ by changing the parity(see solid yellow line with up-triangle and solid purple line with down-triangle in Fig.~\ref{fig2-1}), the correlation function is generally changed into
\begin{eqnarray}
G^{(2)}_8(0){\approx}P_2(\Delta_{21}X_{21})^2/[P^2_1(\Delta_{10}X_{10})^2].
\end{eqnarray}
From the Fig.~\ref{fig2} (c), the fast increase of the output power ${\langle}\hat{X}^-\hat{X}^+{\rangle}/n_c$ dominates the photon distribution, resulting in the two-photon blockade.
It clearly demonstrates the antibunching feature (i.e. $G^{(2)}_8(0)<1$).

By further increasing $\lambda $ to the regime $\lambda{\in} (0.6,0.85)$, the second and third energy levels become nearly degenerate,
which both contribute to the correlation function ${\langle}(\hat{X}^-)^2(\hat{X}^+)^2{\rangle}$.
Moreover, the transition efficient $X_{20}=0$ due to the same parity of $|\phi_2{\rangle}$ and $|\phi_0{\rangle}$.
Hence, the two-photon correlation function is approximately expressed as
\begin{eqnarray}
G^{(2)}_8(0){\approx}\frac{P_2(\Delta_{21}X_{21}\Delta_{10}X_{10})^2+P_3(\Delta_{32}X_{32}\Delta_{21}X_{21})^2}{P^2_1(\Delta_{10}X_{10})^4},
\end{eqnarray}
which can also be verified by the coefficients magnitudes in Fig.~\ref{fig2} (c) and (d).
An antibunching to bunching transition is observed, and the pronounced two-photon signature is exhibited (i.e. $G^{(2)}_8(0){\gg}2$).
The fast decay of ${\langle}\hat{X}^-\hat{X}^+{\rangle}/n_c$ mainly contributes to the enhancement of the $G^{(2)}_8(0)$, generating the giant bunching effect of photons.
This feature is quite distinct from the counterpart in the open Rabi model ($N=1$) in Ref.~\cite{aridolfo2013prl}, where photons are monotonically suppressed by increasing qubit-photon coupling strength.

While in the deep strong coupling regime $\lambda>0.85$, the two-photon correlation function
$G^{(2)}_N(0)$ is dramatically reduced to $2$ due to formation of the thermal state of the Dicke system(see the appendix for the detail)
\begin{eqnarray}
\hat{\rho}_s=\frac{1}{\mathcal{Z}}\sum_m|m{\rangle}_x{\langle}m|e^{-[\omega\hat{A}^\dag_m\hat{A}_m-(\frac{2\lambda{m}}{\sqrt{N\omega}})^2]/(k_BT)},
\end{eqnarray}
with the eigenstate of $\hat{J}_x$ as $\hat{J}_x|m{\rangle}_x=m|m{\rangle}_x$,
the displaced bosonic operator $\hat{A}_m=a+2\lambda{m}/\sqrt{N}$,
and the partition function $\mathcal{Z}=\frac{1}{1-e^{-\omega/k_BT}}\sum_m\exp[-(\frac{2\lambda{m}}{\sqrt{N\omega}})^2/(k_BT)]$.
Hence, the photons are inclined to be classically distributed, which is similar to the counterpart in the Rabi model~\cite{aleboite2016pra}.

\subsection{Effect of finite qubits number}

Next, we  analyze the influence of the finite qubits number on the two-photon correlation function in Fig.~\ref{fig3} (a).
By increasing the qubits number, it is interesting to find that the minimum of the $G^{(2)}_N(0)$ shows monotonic enhancement.
However, the peak of the $G^{(2)}_N(0)$ of the finite size Dicke model is firstly enhanced and then suppressed.
Such optimization can be clearly observed in Fig.~\ref{fig3} (b).
Hence, we conclude that the two-photon correlation can be optimized with finite qubits number.

Moreover, we analyze the scaling behavior of the coupling strength at the extreme value of the $G^{(2)}_N(0)$ with the qubits number $N$ in Fig.~\ref{fig3} (c).
It is found that they behave as
\begin{eqnarray}
[\lambda_{\textrm{max}(\textrm{min})}-\lambda_c]{\propto}N^{-(1{\pm}0.06)},
\end{eqnarray}
where $\lambda_c=\frac{\sqrt{\omega\Delta}}{2}\coth(\frac{\Delta}{4k_BT})$ is the critical coupling strength at finite temperature.
This demonstrates that the $G^{(2)}_N(0)$ may be considered as a potential indicator to detect the criticality of the Dicke model.

\begin{figure}[tbp]
\begin{center}
\includegraphics[scale=0.5]{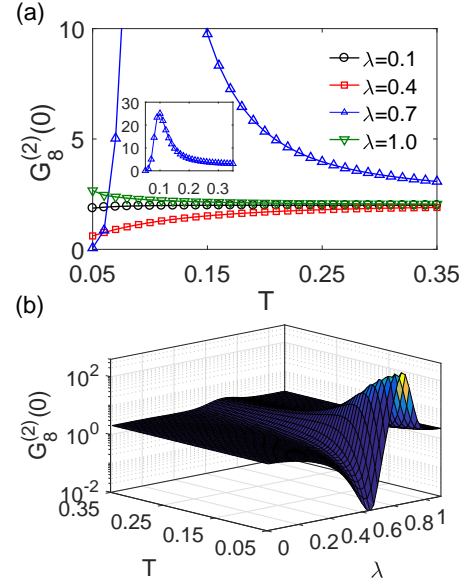}
\end{center}
\caption{(Color online) Two-photon correlation function $G^{(2)}_8(0)$ (a) by tuning temperature $T$ and (b) in a 3D view as a function of $T$ and $\lambda$. The inset shows the complete shape of the two-photon correlation function for $\lambda=0.7$ as a function of the temperature.
The other system parameters are the same as in Fig.~\ref{fig2}.
}~\label{fig4}
\end{figure}

\begin{figure}[tbp]
\begin{center}
\includegraphics[scale=0.4]{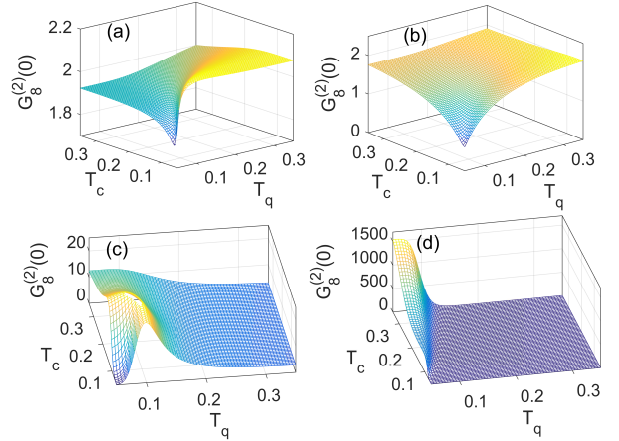}
\end{center}
\caption{(Color online) A 3D view of the two-photon correlation function $G^{(2)}_8(0)$ with
(a) $\lambda=0.1$, (b) $\lambda=0.4$, (c) $\lambda=0.7$, and (d) $\lambda=1.0$.
The other system parameters are the same as in Fig.~\ref{fig2}.
}~\label{fig5}
\end{figure}

\subsection{Effect of finite temperatures of thermal baths}

We investigate the influence of the bath temperatures on the two-photon correlation function in Fig.~\ref{fig4} (a) with finite qubits number (e.g., $N=8$).
In the ultrastrong coupling regime (e.g., $\lambda=0.1$),
by increasing the temperature the two-photon correlation function is enhanced from the anti-bunching to bunching feature,
and approaches thermal distribution ($G^{(2)}_8(0)=2$) in comparatively high temperature regime (e.g., $T=0.35$).
In the qubit-photon coupling regime (e.g., $\lambda=0.7$),
by increasing the temperature a giant two-photon bunching signature is clearly observed.
While in the deep strong coupling regime (e.g., $\lambda=1.0$), the photons are nearly thermally distributed,
with $G^{(2)}_8(0)$ slightly above $2$ in the wide temperature zone.
Hence, we conclude that the optimal coupling strength may enhance the two-photon correlation function.

Then, we give a comprehensive picture of $G^{(2)}_8(0)$ by both tuning temperature and coupling strength in Fig.~\ref{fig4} (b).
It is found that in low temperature regime, the significant signals of the photon blockade and two-photon enhancement are exhibited.
While as the temperature increases, the fluctuation of two-photon correlation function are suppressed monotonically,
finally resulting in  the  thermal state of photons ($G^{(2)}_8(0){\approx}2$).

Next, we investigate the effect of the temperature bias $(T_c{\neq}T_q)$ on the two-photon correlation function in Fig.~\ref{fig5}.
With ultrastrong qubit-photon interaction (e.g., $\lambda=0.1$), the super-thermal behavior of photons (i.e., $G^{(2)}_8(0)>2$) is exhibited with high $T_q$ and low $T_c$ with large temperature bias.
For the coupling case (e.g., $\lambda=0.7$), the giant photon bunching is exhibited with both low $T_q$ and $T_c$.
However, if we further increase the coupling strength (e.g., $\lambda=1.0$), high $T_c$ and low $T_q$ jointly contribute to
the significantly large two-photon bunching.
Hence, we conclude that the two-photon correlation  can be dramatically enhanced with strong qubit-photon interaction and large temperature bias.

\section{Conclusion}

To summarize, we study the zero-time delay two-photon correlation function in the dissipative Dicke model, where the qubits and the photons are
individually coupled to thermal baths, respectively.
The quantum dressed master equation is applied to analyze the steady state behavior of the Dicke system with strong qubit-photon interaction.
We investigate the influence of the qubit-photon coupling strength in the  two-photon correlation function.
An anti-bunching to bunching transition and giant two-photon correlation are clearly exhibited in the ultrastrong coupling regime.
We also analyze the effect of the finite qubits number on the two-photon correlation function.
It is found that the maximal two-photon bunching feature is observed with the optimal qubits number.
Moreover, the coupling strengthes at the extreme values of two-photon correlation function scale
as $[\lambda_{\textrm{max}(\textrm{min})}-\lambda_c]{\propto}1/N$ with $\lambda_c$ the superradiant phase transition of the Dicke model at finite temperature.
Then, we analyze the effect of the finite temperature on the two-photon correlation.
The low bath temperature is crucial to exhibit the two-photon blockade and bunching behaviors.
We also study the two-photon correlation function with temperature difference of thermal baths.
It is found that strong qubit-photon interaction and large temperature bias jointly contribute to the giant two-photon bunching.

Finally, we should note that the finite-time delay two-photo correlation function is also a powerful tool to analyze the photon distribution, e.g., photon blockade in optomechanics~\cite{prabl2011prl}.
We may apply the finite-time delay correlation function in further to analyze the photon behavior of the dissipative Dicke model.

\section{Acknowledgement}

W.C. is supported by the National Natural Science Foundation of China under Grant No. 11704093
and the Opening Project of Shanghai Key Laboratory of Special Artificial Microstructure Materials and Technology.
G.X. and X.H. acknowledge support from NSFC under
Grants No. 11835011 and No. 11774316.

\appendix

\section{Two-photon correlation function at strong qubit-photon coupling}
In the strong qubit-photon coupling limit, the qubit tunneling is strongly dressed,
and the Hamiltonian at Eq.~(\ref{h0}) is simplified
$\hat{H}_{strong}{\approx}\omega\hat{a}^\dag\hat{a}+\frac{2\lambda}{\sqrt{N}}(\hat{a}^{\dag}+\hat{a})\hat{J}_x$,
which can be re-expressed as
\begin{eqnarray}
\hat{H}_{strong}{\approx}\sum_{m}|m{\rangle}_x{\langle}m|[\omega\hat{a}^\dag\hat{a}+\frac{2\lambda{m}}{\sqrt{N}}(\hat{a}^{\dag}+\hat{a})],
\end{eqnarray}
where $\hat{J}_x|m{\rangle}_x=m|m{\rangle}_x$.
If we define the displaced bosonic operator associated with the angular momentum as
$\hat{A}_m=\hat{a}+\frac{2\lambda{m}}{\sqrt{N}}$, the Hamiltonian is given by
\begin{eqnarray}
\hat{H}_{strong}{\approx}\sum_{m}|m{\rangle}_x{\langle}m|[\omega\hat{A}^\dag_m\hat{A}_m-(\frac{2\lambda{m}}{\sqrt{N\omega}})^2].
\end{eqnarray}
Hence, the steady state thermal state is given by
\begin{eqnarray}
\hat{\rho}_s=\frac{1}{\mathcal{Z}}\sum_md_m|m{\rangle}_x{\langle}m|e^{-\omega\hat{A}^\dag_m\hat{A}_m/(k_BT)},
\end{eqnarray}
where the temperature $T_q=T_c=T$,
$d_m=\exp{[(\frac{2\lambda{m}}{\sqrt{N\omega}})^2/k_BT]}$,
and $\mathcal{Z}$ is the partition function to normalize $\hat{\rho}_s$.
Moreover, the photon detection operator is specified as
\begin{eqnarray}
\hat{X}^-=-i\omega\sum_m|m{\rangle}_x{\langle}m|\hat{A}^{\dag}_m.
\end{eqnarray}
Therefore, it is easy to calculate the correlation functions at thermal state as
\begin{eqnarray}
{\langle}\hat{X}^-\hat{X}^+{\rangle}&=&\omega/[e^{\omega/k_BT}-1],\\
{\langle}(\hat{X}^-)^2(\hat{X}^+)^2{\rangle}&=&2\omega^2/[e^{\omega/k_BT}-1]^2.
\end{eqnarray}
Finally, we obtain the two-photon correlation function as $G^{(2)}_N(0)=2$.



\end{document}